\documentclass[aps,prl,reprint,groupedaddress,longbibliography]{revtex4-1}

\usepackage{graphicx}
\usepackage{dcolumn}
\usepackage{bm}
\usepackage{amsmath}
\usepackage{color}
\usepackage[usenames,dvipsnames,svgnames,table]{xcolor}
\usepackage[colorlinks=true,
            linkcolor=blue,
            urlcolor=blue,
            citecolor=blue]{hyperref}
\usepackage{siunitx}
\DeclareSIUnit\gauss{G}

\newcommand{\fig}[5][t]{\begin{figure}[#1]\includegraphics[width=86mm]{#3.pdf}\vspace{-0.25 cm}\caption{#4}\label{fig:#5}\end{figure}}
\newcommand{\figx}[5][t]{\begin{figure*}[#1]\includegraphics[width=172mm]{#3.pdf}\vspace{-0.25 cm}\caption{#4}\label{fig:#5}\end{figure*}}

\newcommand{\figOneFile}[0]{Figure1}        
\newcommand{\figTwoFile}[0]{Figure2}        
\newcommand{\figThreeFile}[0]{Figure3}      
\newcommand{\figFourFile}[0]{Figure4}    	
\newcommand{\figFiveFile}[0]{Figure5}     	
\newcommand{\sfigZeroFile}[0]{FigureS0}     
\newcommand{\sfigOneFile}[0]{FigureS1}      
\newcommand{\sfigTwoFile}[0]{FigureS2}      
\newcommand{\sfigThreeFile}[0]{FigureS3}    
\newcommand{\sfigFourFile}[0]{FigureS4}     

\begin{document}

\title{Two-Photon Blockade in an Atom-Driven Cavity QED System}

\author{Christoph Hamsen}
\email[]{Christoph.Hamsen@mpq.mpg.de}
\author{Karl Nicolas Tolazzi}
\author{Tatjana Wilk}
\author{Gerhard Rempe}
\affiliation{Max-Planck-Institut f\"{u}r Quantenoptik, Hans-Kopfermann-Str. 1, 85748 Garching, Germany}

\date{\today}

\begin{abstract}
    Photon blockade is a dynamical quantum-nonlinear effect in driven systems with an anharmonic energy ladder. For a single atom strongly coupled to an optical cavity, we show that atom driving  gives a decisively larger optical nonlinearity than cavity driving. This enhances single-photon blockade and allows for the implementation of two-photon blockade where the absorption of two photons suppresses the absorption of further photons. As a signature, we report on three-photon antibunching with simultaneous two-photon bunching observed in the light emitted from the cavity. Our experiment constitutes a significant step towards multi-photon quantum-nonlinear optics.
\end{abstract}

\maketitle
An open driven quantum system exhibits fluctuations that reflect its walk through Hilbert space. Blocking parts of the Hilbert space can reduce these fluctuations and stabilize the output. For discrete variables like particle number, blockade occurs for sufficiently strong interaction between the involved quanta. Examples include the Coulomb force for electrons or the effective interaction between photons in an optically nonlinear medium. The latter has been used to realize single-photon blockade~\cite{Imamoglu1997} where $n=1$ photon blocks further photons so that they are emitted one by one~\cite{Kimble1977, Birnbaum2005, Dayan2008, Faraon2008a, Reinhard2011, Lang2011, Hoffman2011}. The challenge now is to scale the blockade to $n>1$ photons~\cite{Shamailov2010, Miranowicz2013, Carmichael2015} and produce a photon stream with at most $n$ photons. Such quantum scissors could lead to novel applications in multiphoton quantum-nonlinear optics like an $n$-photon source~\cite{Chang2014}.

An ideal platform for the implementation of an optical $n$-photon blockade is cavity quantum electrodynamics (QED) which strongly couples a single two-level atom, perfectly blockaded at one photon, to a cavity that is completely unblocked. Both subsystems alone fail to show multi-photon blockade: the cavity needs the nonlinearity introduced by the atom, and the atom needs access to the larger Hilbert space provided by the cavity. Only the combined system with its anharmonic energy-level structure provides the necessary photon-number dependent nonlinearity. Nevertheless, realization of multi-photon blockade is challenging due to the limited atom-cavity coupling strength that has so far been obtained~\cite{Vahala2003, Devoret2007, Schuster2008}. Although strategies have been proposed to improve the blockade by extension to a three- or four-level atom involving electromagnetically induced transparency~\cite{Rebic1999, Souza2013a} or Raman scattering~\cite{Rosenblum2011, Rosenblum2015a}, multi-photon blockade has not been observed in optical systems. Its demonstration in circuit QED seems pending, too, although well-resolved multi-photon transitions have been investigated~\cite{Fink2008c, Deppe2008, Bishop2009}.

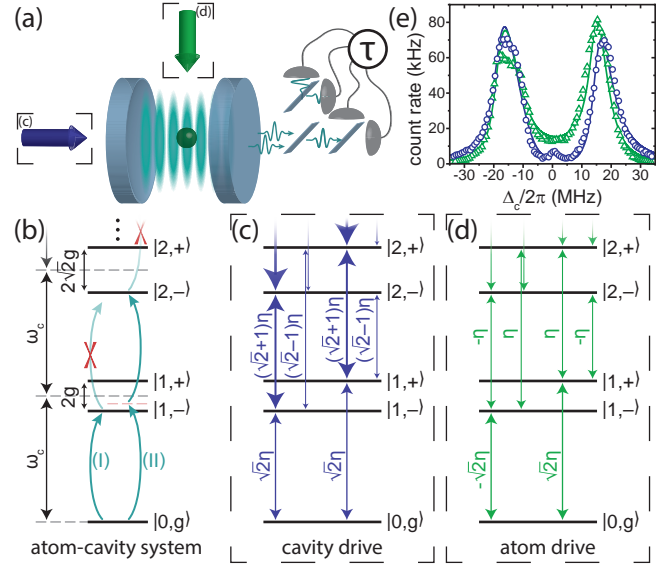
\fig{1}{\figOneFile}{(Color online) Sketch of the experimental setup and physical system. As depicted in (a), a single atom is trapped at the antinode of an intracavity light field. The anharmonic energy-ladder system (b) can either be excited via a cavity (blue, effective driving strengths see (c)) or atom drive (green, effective driving strengths see (d)). The resulting cavity field is then monitored via an extended Hanbury Brown and Twiss detection setup. Spectra for driving the cavity (blue circles) or the atom (green triangles) are shown in (e). The thick colored lines are fits of a model considering residual thermal excitation and possible remnants of the empty cavity~\cite{supple}. Symbols: cavity frequency $\omega_c$, coupling strength $g$, driving strength $\eta$, probe-cavity detuning $\Delta_c$.}{fig1}
%
This Letter reports on the first experimental observation of two-photon blockade with a strongly coupled atom-cavity system. Specifically, we demonstrate an increased excitation of the system's second energy manifold in combination with a suppressed excitation of the third and higher manifolds. As a signature, the light emitted from the cavity exhibits a pronounced three-photon antibunching with simultaneous two-photon bunching when driving the system close to a two-photon resonance~\cite{supple}. We show that two-photon blockade exists only for excitation of the system via the atom, while cavity driving at this frequency yields strong bunching of second- and third-order photon correlations. The novel dependence on the excitation path can be understood intuitively as a consequence of bosonic enhancement of photons when driving the cavity, an effect which facilitates climbing up the ladder of dressed atom-cavity states. The atom, in contrast, can absorb only one photon at a time and thus makes the unwanted climbing more difficult. We therefore claim that in order to exploit the full optical nonlinearity of the system for the realization of, e.g., an $n$-photon absorber or an $n$-photon emitter, it is more favorable to drive the atom instead of the cavity.

The dependence of the nonlinear behavior on the driven component can be expressed quantitatively by calculating the transition strengths in the dressed state basis. The driven atom-cavity system as depicted in Fig.\,\ref{fig:fig1}(a) is well described by the Jaynes-Cummings Hamiltonian~\cite{Jaynes1963} plus a driving term $H_d$, here in rotating-wave approximation:
\begin{equation}\label{Eq.:JC-Hamiltonian}
H=\hbar \Delta_{a}\, \hat{\sigma}^\dagger\hat{\sigma}+\hbar \Delta_{c}\, \hat{a}^\dagger \hat{a} + \hbar g\,(\hat{a}^\dagger \hat{\sigma}+\hat{\sigma}^\dagger \hat{a}) + H_d
\end{equation}
where $\Delta_a=\omega_d-\omega_a$ ($\Delta_c=\omega_d-\omega_c$) is the atom (cavity) detuning with respect to the driving frequency $\omega_d$, $\hat{\sigma}^\dagger$ ($\hat{\sigma}$) is the atomic raising (lowering) operator, and $\hat{a}^\dagger$ ($\hat{a}$) is the photon creation (annihilation) operator with $\hat{n}=\hat{a}^\dagger \hat{a}$ being the photon number operator. While the first two terms in Eq.\,\ref{Eq.:JC-Hamiltonian} correspond to the bare energy eigenstates of emitter and resonator, the third term describes their interaction with coupling strength $g$. This yields energy eigenstates that form an anharmonic ladder of doublets ($|n,\pm\rangle=(|n,g\rangle\pm|n-1,e\rangle)/\sqrt{2}$) split by $2\sqrt{n}g$, referred to as dressed states (Fig.\,\ref{fig:fig1}(b)). Single- and two-photon blockade are then expected for resonant one- and two-photon excitation of the first (I) and second (II) manifold, respectively.

The last term in Eq.\,\ref{Eq.:JC-Hamiltonian} describes the excitation via the driving field. The energy structure remains unaffected as long as the drive strength is much smaller than $g$ and does not exceed the atomic polarization decay rate $\gamma$ and cavity-field decay rate $\kappa$~\cite{Alsing1992}. However, the corresponding excitation strengths between different manifolds differ whether the cavity is driven, $H_d=\hbar\eta_{\text{\it{c}}}\,(\hat{a}+\hat{a}^\dagger)$, or the atom, $H_d=\hbar\eta_{\text{\it{a}}}\,(\hat{\sigma}+\hat{\sigma}^\dagger)$~\cite{Alsing1991}. Here, $\eta_{\text{\it{c}}}$ ($\eta_{\text{\it{a}}}$) is the strength of the cavity (atom) drive. Both strengths are expressed for the bare eigenstates of the system without atom-cavity interaction. Reformulation in the dressed state basis of the coupled system ($|n,\pm\rangle$) yields effective strengths $\widetilde{\eta}_{\text{\it{c}}}/2$ or $\widetilde{\eta}_{\text{\it{a}}}/2$ for cavity or atom drive, respectively. For the transition from the ground state to the first manifold, $|0,g\rangle\rightarrow|1,\pm\rangle$, these are $\widetilde{\eta}_{\text{\it{a}}}=\pm\sqrt{2}\eta_{\text{\it{a}}}$ and $\widetilde{\eta}_{\text{\it{c}}}=\sqrt{2}\eta_{\text{\it{c}}}$. For the transition from the $n$th to the $(n+1)$th manifold, and in case of cavity driving, bosonic bunching causes symmetry conserving transitions, ($|n,\pm\rangle\rightarrow|n+1,\pm\rangle$), to be strongly enhanced by $\widetilde{\eta}_{\text{\it{c}}}=(\sqrt{n+1}+\sqrt{n})\eta_{\text{\it{c}}}$ whereas those that change symmetry, ($|n,\pm\rangle\rightarrow|n+1,\mp\rangle$), are suppressed, $\widetilde{\eta}_{\text{\it{c}}}=(\sqrt{n+1}-\sqrt{n})\eta_{\text{\it{c}}}$ (Fig.\,\ref{fig:fig1}(c)). For an atom drive, all transitions have equal strengths, with the sign being that of the upper state, $\widetilde{\eta}_{\text{\it{a}}}=\pm\eta_{\text{\it{a}}}$ (Fig.\,\ref{fig:fig1}(d)).

As a consequence, resonant driving of the $n$th manifold via the cavity reduces the suppression of higher excitations since the corresponding transition strengths increase. In contrast, the transition strengths remain constant when driving the atom. As will be shown in the following, the resulting stronger suppression of higher rungs for atom excitation manifests itself in an improved purity of single-photon emission on the first manifold and enables two-photon blockade on the second manifold.

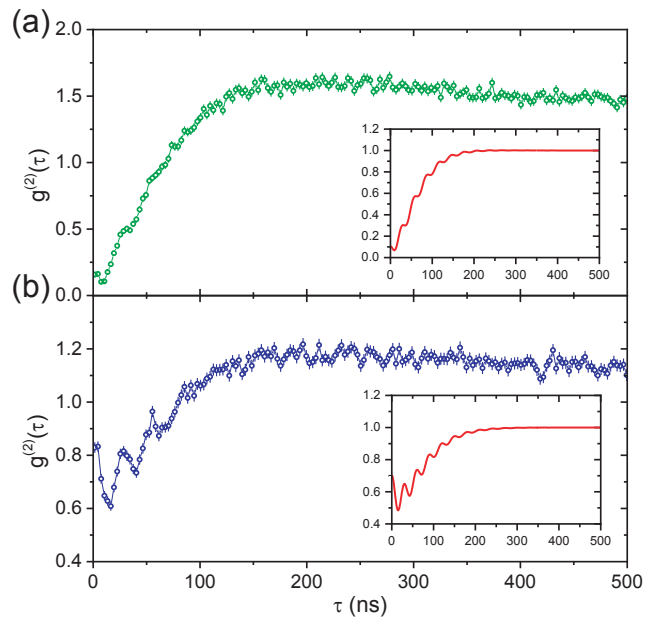
\fig{2}{\figTwoFile}{(Color online) The second-order photon correlation function for (a) atom and (b) cavity excitation of the first manifold at $\Delta_{c}/2\pi=\SI{+18}{\mega\hertz}$ with a binning of \SI{3}{\nano\second} reveals single-photon blockade. Here and in all following figures, error bars are statistical indicating one standard deviation, and theory (insets) is calculated via numerical solution of the master equation~\cite{supple}. Theory is shown for qualitative comparison, deviations to experimental results stem from atomic motion and position distribution of atoms within the cavity mode. }{fig2}
%
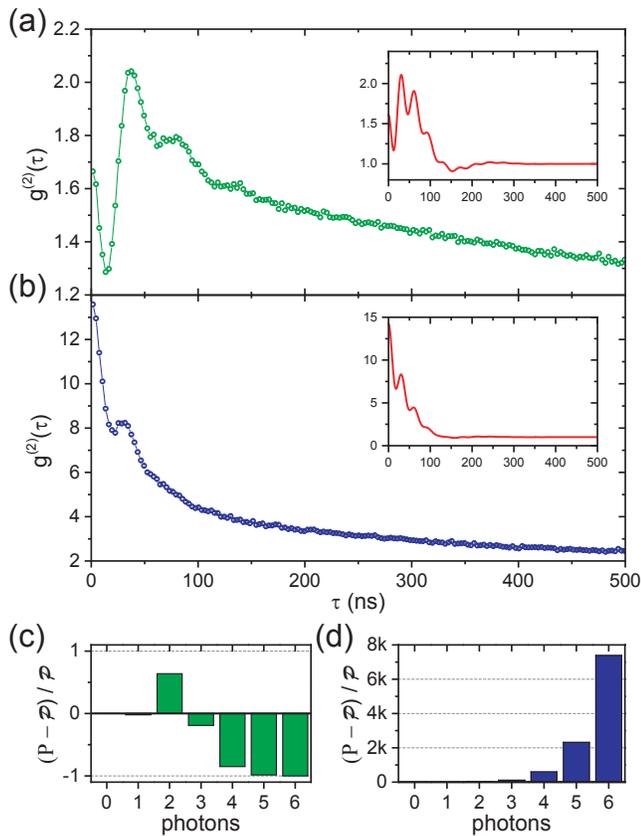
\fig{3}{\figThreeFile}{(Color online) Close to the second manifold at $\Delta_{c}/2\pi=\SI{+9}{\mega\hertz}$, photon correlations with \SI{3}{\nano\second} binning for (a) atom and (b) cavity driving show bunching. Insets depict the corresponding theory, which we add for qualitative comparison. The relative deviations of the simulated photon distribution to a Poisson distribution of the same mean photon number indicates two-photon blockade for atom (c) but not cavity (d) excitation. In (d), the ordinate is scaled by a factor of k=1000.}{fig3}
%
In our system, a single $^{87}Rb$ atom ($\gamma/2\pi=\SI{3.0}{\mega\hertz}$) is loaded into the center of a high-finesse Fabry-Perot resonator with length \SI{200}{\micro\metre} and a field decay rate $\kappa/2\pi=\SI{2.0}{\mega\hertz}$~\cite{supple}. Two blue- and one red-detuned standing-wave optical dipole traps form a three-dimensional lattice that confines the atom to an antinode of the cavity field~\cite{Reiserer2013c}. The dynamical Stark shift, mainly caused by the red-detuned \SI{800}{\nano\metre} trap, reduces the atom-cavity detuning, $\Delta_{ac}/2\pi=(\omega_a-\omega_c)/2\pi=\SI{-15.2}{\mega\hertz}$, to the $F=2\leftrightarrow F'=3$ transition of the $D_{2}$ line at \SI{780}{\nano\metre} to only a few \si{\mega\hertz}.
We use the transition with the largest dipole matrix element between Zeeman states $m_F=+2\,\leftrightarrow\,m_F'=+3$.
Here, an atom-cavity coupling strength of $g/2\pi=\SI{20}{\mega\hertz}$ puts the experiment well into the strong-coupling regime of cavity QED, $g\gg(\kappa,\gamma)$.

As long as the atom is trapped (typically \SI{5}{\second}), we repeat our measurement sequence with a rate of \SI{2}{\kilo\hertz}. This sequence consists of a cooling interval, state preparation of the $F=2,m_F=+2$ state and a probe interval during which we apply the respective probe (alternating from sequence to sequence) at the desired frequency and record the transmitted signal on 4 single-photon detectors with a timing resolution of \SI{1}{\nano\second}.The power is chosen such that we remain in the weak driving regime, $\eta_{a,c}\leq(\kappa,\gamma)$.

Spectra for atom and cavity driving are depicted in Fig.\,\ref{fig:fig1}(e). In both cases, the distinct splitting of the normal modes reflects the strong coupling of the system. We deduce an experimental coupling constant of $g/2\pi=\SI{16.38+-0.04}{\mega\hertz}$. The stronger drop of transmission in case of cavity driving results from the atomic antiresonance caused by destructive interference when exciting the cavity~\cite{Sames2014}. This also slightly increases the observed normal mode splitting.

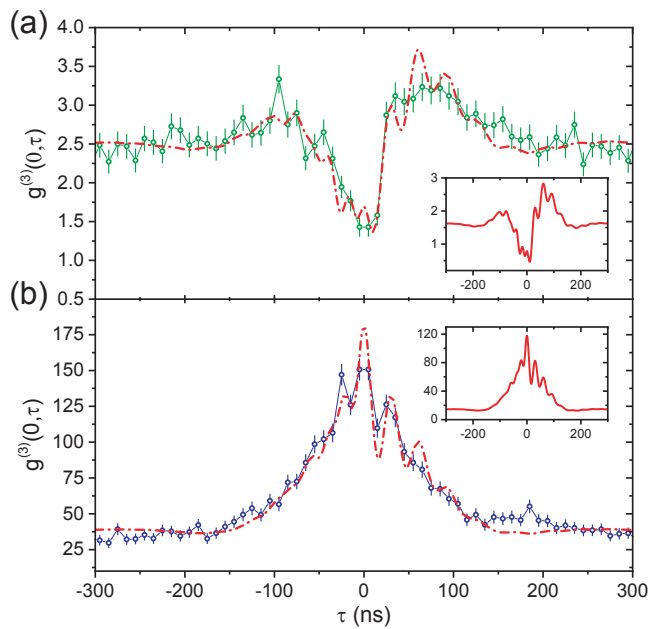
\fig{4}{\figFourFile}{(Color online) The third-order photon correlation function $g^{(3)}(0,\tau)$ is depicted for (a) atom and (b) cavity excitation close to the second manifold with a binning of \SI{10}{\nano\second}. For qualitative comparison, the theory for the same parameters (inset) has been scaled and shifted to fit experimental data (dash dotted red lines).}{fig4}
%
To demonstrate single-photon blockade and its dependence on the driven component, we start by exciting the system close to the first manifold. The measured second-order photon correlation function $g^{(2)}(\tau)=\langle\hat{n}\cdot\hat{n}(\tau)\rangle/\langle \hat{n}\rangle^2$ (normal and time ordered) for atom driving ($\eta_a/2\pi\approx\SI{0.55}{\mega\hertz}$) is shown in Fig.\,\ref{fig:fig2}(a) with the corresponding theory as an inset. A strong sub-Poissonian antibunching with a $g^{(2)}(0)=\SI{0.16+-0.01}{}$ and a rising slope indicate emission of single light quanta due to a strong blockade of multiple excitations (compare Fig.\,\ref{fig:fig1}(b,I)). We observe a small and rapid oscillation at approximately twice the coupling rate $g$ known as vacuum Rabi oscillation~\cite{Rempe1991}. It originates from the coherent energy exchange between atom and cavity. We estimate the coupling rate from the second oscillation maximum at \SI{31.5+-1.5}{\nano\second} to be \SI{15.9+-0.8}{\mega\hertz} which is in good agreement with the fitted value from the spectrum. The non-classical behavior disappears on a timescale determined by the decay rate of the excited dressed state $(\frac{\kappa+\gamma}{2})^{-1}\approx\SI{64}{\nano\second}$. The classical value, achieved for large correlation times, deviates from \SI{1}{} due to motion and residual displacement from the cavity-mode center~\cite{Rempe1991, supple}.

Excitation of the cavity ($\eta_{c}/2\pi\approx\SI{0.55}{\mega\hertz}$) on the first manifold is depicted in Fig.\,\ref{fig:fig2}(b) and yields qualitatively the same behavior. However, the value $g^{(2)}(0)=\SI{0.83+-0.02}{}$ is much larger, and stronger vacuum Rabi oscillations indicate significant excitation of higher manifolds. In accordance with theory, atom driving does exhibit a far stronger photon blockade effect despite the same energy-level structure.

In order to investigate two-photon blockade, we tune the drives close to the second manifold (compare Fig.\,\ref{fig:fig1}(b,II)) and increase their strengths to $\eta_a/2\pi\approx\SI{1.6}{\mega\hertz}$ and $\eta_{c}/2\pi\approx\SI{1.1}{\mega\hertz}$, approaching the cavity decay rate to allow for significant population of higher states without yet affecting the level structure. As shown in Fig.\,\ref{fig:fig3}(a) and (b), this yields super-Poissonian emission in both cases since $g^{(2)}(0)>1$ which is indicative of higher photon numbers. While cavity excitation shows the expected bunching behavior~\cite{Kubanek2008}, the observed dynamics for atom driving is more complex. The interplay between conflicting mechanisms, a two-photon resonance on one hand, and an emitter that can only absorb one excitation at a time on the other hand, leads to a novel photon-concatenation effect. Since the rate of coherent energy exchange between atom and cavity exceeds the spontaneous decay rate of the system, higher manifolds are populated in stepwise excitation via the emitter. As a consequence, we observe that the second-order correlation function peaks \SI{37.5+-1.5}{\nano\second} after the trigger photon which indicates that the coupling rate rather than the lifetime determines the probability for detection of a second photon, in contrast to the first-manifold dynamics described above. As $g^{(2)}(\tau)>g^{(2)}(0)$, this behavior violates the Cauchy-Schwarz inequality and is thus quantum in nature~\cite{Mandel1995,Mielke1998}.

While Fig.\,\ref{fig:fig3}(a) and (b) indicate multi-photon emission, a two-photon blockade furthermore requires suppression of excitation to even higher manifolds. To illustrate this, we calculate the full photon-number distribution $\text{P(n)}$ and compare this to a Poisson distribution $\mathcal{P}\text{(n)}$ of the same mean photon number as depicted in Fig.\,\ref{fig:fig3}(c) and (d). For cavity driving, the relative population grows with the excitation number as expected due to bosonic enhancement. In case of atom excitation, we see enhanced two-photon emission while higher Fock states are increasingly suppressed. The latter condition can be understood as truncation of the Hilbert space and indicates two-photon blockade that for our parameters is only visible for atom driving.

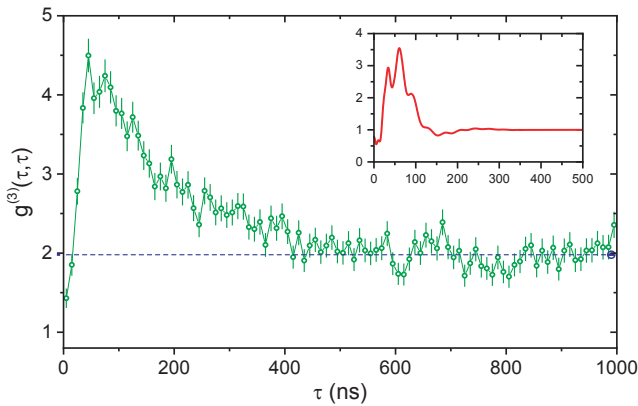
\fig{5}{\figFiveFile}{(Color online) The third-order photon correlation function $g^{(3)}(\tau,\tau)$ for atom driving close to the second manifold with a binning of \SI{10}{\nano\second} is depicted. The inset shows the result expected by theory. The blue empty marker and dashed line indicate the long-time average for \SI{1}{\micro\second} to \SI{4}{\micro\second} (after the correlation has settled) which is clearly above $g^{(3)}(0,0)$.}{fig5}

For demonstrating two-photon blockade, we evaluate the third-order photon correlation $g^{(3)}(\tau_1,\tau_2)=\langle\hat{n}\cdot\hat{n}(\tau_1)\cdot\hat{n}(\tau_1+\tau_2)\rangle/\langle\hat{n}\rangle^3$ (normal and time ordered). Here, we discuss two specific cases. We start with the dynamically interesting case of $(\tau_1,\tau_2)=(0,\tau)$ where the third-order correlation yields information on the conditional evolution of $\langle\hat{n}\rangle$ ($\langle\hat{n}^2\rangle$), i.e. the dynamics on the first (second) manifold for positive (negative) $\tau$~\cite{Koch2011}. This is depicted in Fig.\,\ref{fig:fig4}(a) and (b) for atom and cavity driving, respectively. We find good qualitative agreement with theory that for comparison has been shifted and scaled to again compensate for effects due to atomic motion and residual position distribution within the cavity. Note that for large $\tau$ one cannot expect $g^{(3)}(0,\tau)$ to approach 1, but the value of $g^{(2)}(0)\cdot g^{(2)}(\tau)$ since two of the photons are correlated for any $\tau$~\cite{Koch2011a}. The asymmetry and different oscillation frequencies for positive and negative times reflect the coherent evolution on the first and second manifold at frequency $2g$ and $2\sqrt{2}g$, respectively~\cite{Koch2011}. Most striking is that atom and cavity excitation exhibit very different behavior towards zero time delay: We observe antibunching when driving the atom in contrast to a strong bunching for cavity excitation. As a consequence, higher photon numbers are suppressed (enhanced) when exciting the atom (cavity) into the second manifold.

To prove suppression of three-photon emission, we evaluate $g^{(3)}(\tau,\tau)$ for atom driving as shown in Fig.\,\ref{fig:fig5}. For time intervals exceeding the time scale of the internal coherence, $\tau\gg2/(\kappa+\gamma)$, $g^{(3)}(\tau,\tau)$ is proportional to the probability of detecting three uncorrelated photons. In contrast to theory, $g^{(3)}(0,0)=\SI{1.43+-0.12}{}$ is above \SI{1}{}, the value expected for a Poissonian light field. However, we do significantly underpass the long-term averaged value of \SI{1.98+-0.01}{} that serves as a reference for uncorrelated photons. This value is above \SI{1}{} due to technical fluctuations that shift $g^{(3)}(\tau,\tau)$ to higher values~\cite{supple}. To confirm this, we calculate our photon distribution $P(n)$ from the number of photons per measurement interval, averaged over many realizations, and deduce a value of $g^{(3)}(0,0)=\frac{\sum_{n}n(n-1)(n-2)P(n)}{(\sum_{n}nP(n))^3}=\SI{1.99}{}$ for uncorrelated photons. This value agrees very well with the long-time averaged $g^{(3)}(\tau,\tau)$ and proves an increased variance of the field, likely due to residual atom motion and a distribution of positions with respect to the cavity mode and atom drive~\cite{supple}. We conclude that the $g^{(3)}(\tau,\tau)$ therefore demonstrates a two-photon blockade where the probability of detecting more than two photons for zero time delay is reduced.

In conclusion, we have shown that driving the quantum emitter instead of the resonator improves the nonlinear response of the strongly coupled system. This allows us to demonstrate both single- and two-photon blockade. Future experiments could explore the extension of the blockade mechanism to even higher photon numbers. For example, simulations indicate that three-photon blockade seems feasible with our system. As blockade truncates the high end of the photon-number distribution, any additional reduction of the low end~\cite{Faraon2008a, Kubanek2008} may enable carving of various non-classical photon states like those containing $n$ photons. Direct production of $n$-photon states has also been proposed for strong atom driving, $\eta_{\text{\it{a}}}\gg g$, with the cavity tuned as to selectively enhance a specific $n$-photon transition between dressed atom-laser states~\cite{Munoz2014}. Selective population of higher-energetic atom-cavity states might be possible by stepwise excitation of the symmetry-changing transitions ($|n,\pm\rangle\rightarrow|n+1,\mp\rangle$). When exciting the atom instead of the cavity, these transitions exhibit larger and thus more favorable strengths~\cite{Fink2008c}. Finally, driving atom and cavity simultaneously might enable a quantum interference induced photon blockade where single-photon emission results from destructive interference between different transition paths~\cite{Tang2015a}.

\begin{acknowledgments}
We thank H. Chibani, B. Dayan, A. Gonz\'{a}lez-Tudela and S. D{\"{u}}rr for discussions. P. A. Altin, M. Bernard-Schwarz and A. C. Eckl contributed to the implementation of the experiment. C.H. acknowledges support from the Deutsche Forschungsgemeinschaft via the excellence cluster Nanosystems Initiative Munich (NIM).
\end{acknowledgments}

\bibliographystyle{apsrev4-1}

\clearpage

\setcounter{equation}{0}
\setcounter{figure}{0}
\setcounter{table}{0}
\setcounter{page}{1}
\makeatletter
\renewcommand{\theequation}{S\arabic{equation}}
\renewcommand{\thefigure}{S\arabic{figure}}
\renewcommand{\bibnumfmt}[1]{[S#1]}
\renewcommand{\citenumfont}[1]{#1}

\part*{\Large\centering Supplementary Information}
\section*{Multi-Photon Blockade}
An open quantum system shows $n$-photon blockade if its Hilbert space is restricted to states containing at most $n$ quanta. Such a system must meet the following requirements: First, it must be able to store $n$ excitations at the same time. Second, for the truncation of the Hilbert space it requires some kind of nonlinearity to suppress the probability for $n+1$ photons. Moreover, for proper characterization of the system, it must posses an output channel that allows for simultaneous emission of all stored excitations. In case of a single-atom cavity-QED system, it is usually unimportant which component (atom or cavity) is driven. However, only the cavity can emit multiple excitations at the same time, and therefore the light field emitted from the cavity (not the atom) must be used for characterization.

For now, it is sufficient to treat the system as a 'black box' with at least one input that is driven with a coherent field and one spatio-temporal output mode that is used to characterize the system~\cite{Birnbaum2005a}.
In case of an ideal $n$-photon blockade, the light field emitted from the system shows the following photon number distribution~\cite{Miranowicz2013}
\begin{subequations}
	\begin{align}
	(i)\quad  &   P(m)=  0 \text{ for } m>n \label{Eq:PerfectPStat_a} \\
	(ii)\quad &  P(n)\neq  0  \label{Eq:PerfectPStat_b}
	\end{align}
\end{subequations}
with normalization $\sum_{m=0}^{\infty} {P(m)}=1$.
While the first condition~\ref{Eq:PerfectPStat_a} reflects the fact that at most $n$ photons are emitted at the same time, the second condition~\ref{Eq:PerfectPStat_b} is set to exclude that the system is already ($n-1$)-photon blockaded.
These two conditions can be translated into conditions for the normalized equal-time $k$th-order photon correlation
\begin{align}
g^{(k)}=\sum\limits_{m=k}^{\infty} \frac{m!}{(m-k)!} \frac{P(m)}{\langle\hat{m}\rangle^k}
\end{align}
with $\langle\hat{m}\rangle$ the mean photon number of the output mode. Photon-correlation functions are preferred over a direct measurement of the photon number distribution, since  the physically interesting effects are not prone to attenuation due to low detection efficiencies.
In order to verify if a system is $n$-photon blockaded, the emitted light field has to fulfill the following conditions
\begin{subequations}
	\begin{align}\label{Eq:PerfectBlockade_a}
	(i)\quad  & g^{(n+1)}  =  0 \\ \label{Eq:PerfectBlockade_b}
	(ii)\quad & g^{(n)} \neq  0.
	\end{align}
\end{subequations}
These strict conditions can only be fulfilled with a perfectly blockaded system, which is hard to achieve in an experiment.
If the $n$-photon blockade does not work perfectly, we expect that $P(m)\neq 0$ even for $m>n$, but that the probability to get more than $n$ photons is suppressed with respect to a Poisson distribution with $\mathcal{P}\text{(m)}=\langle\hat m\rangle^m e^{-\langle\hat m\rangle}/m!$ which has the same average photon number $\langle\hat m\rangle$ as the emitted light.
Such a comparison makes sense since perfectly uncorrelated photons follow a Poisson distribution.
In this case the two conditions  Eqs.~\ref{Eq:PerfectPStat_a} and \ref{Eq:PerfectPStat_b} can be transferred into
\begin{subequations}
	\begin{align}\label{Eq:ImPerfectBlockade_a}
	(i)\quad  & P(m)  < \mathcal{P}(m) \text{ for } m>n \\ \label{Eq:ImPerfectBlockade_b}
	(ii)\quad & P(n)  \geq \mathcal{P}(n).
	\end{align}
\end{subequations}
Again, the first condition ensures that all photon numbers above $n$ are suppressed, and the second condition ensures that this is not already the case for $n-1$ photons.
In the following we consider the case of a small mean photon number, $\langle\hat{m}\rangle\ll 1$, and a photon number distribution that fulfills the condition $P(m)\gg P(m+1)$.
These conditions are typically fulfilled in cavity QED experiments that have a strong nonlinearity and are therefore inherently hard to excite.
Then, it is sufficient to show that Eq.~\ref{Eq:ImPerfectBlockade_a} holds for $n+1$, whereas $P(n+1)$ can be approximated using the correlation function $g^{(n+1)}$ (as $P(m)$ can be neglected for all $m>n+1$).
The condition~\ref{Eq:ImPerfectBlockade_a} then reads
\begin{align}
\begin{split}
P(n+1) = \frac{\langle\hat m\rangle^{(n+1)}} {(n+1)!} \cdot g^{(n+1)}  \\
<   \mathcal{P}(n+1) = \frac{\langle\hat m\rangle^{(n+1)}} {(n+1)!} \cdot e^{-\langle\hat m\rangle},
\end{split}
\end{align}
which can be simplified using the Taylor expansion for $e^{-\langle\hat m\rangle}$ to
\begin{align}
g^{(n+1)} & < 1-\frac{\langle\hat m\rangle}{1!} + \frac{\langle\hat m\rangle^2}{2!} -\frac{\langle\hat m\rangle^3}{3!} +\text{...}
\end{align}
For $\langle\hat{m}\rangle\ll 1$ the lower bound of the right hand side of the inequality is $1-\langle\hat m\rangle$, so that the condition formulated in Eq.~\ref{Eq:ImPerfectBlockade_a} is always fulfilled if the following inequality holds
\begin{align}\label{Ineq:ImperfectPB_a}
g^{(n+1)} < 1-\langle\hat m\rangle.
\end{align}

For the condition formulated in Eq.~\ref{Eq:ImPerfectBlockade_b} we can also find a reformulation in terms of correlation functions under the same restrictions as Eq.~\ref{Ineq:ImperfectPB_a} was formulated.
\begin{align}
\begin{split}
P(n) = \frac{\langle\hat m\rangle^n}{n!} \cdot g^{(n)}  - \frac{\langle\hat m\rangle^{(n+1)}}{n!} \cdot g^{(n+1)}\\
\geq   \mathcal{P}(n) = \frac{\langle\hat m\rangle^n} {n!} \cdot e^{-\langle\hat m\rangle},
\end{split}
\end{align}
which reduces to
\begin{align}\label{Ineq:ImperfectPB_b}
g^{(n)} \geq 1-\frac{\langle\hat m\rangle^2}{2}
\end{align}
when inserting the upper limit for $g^{(n+1)}=1-\langle\hat m\rangle$ from Eq.~\ref{Ineq:ImperfectPB_a} into the inequality. Again, if this stricter condition is met, Eq.~\ref{Eq:ImPerfectBlockade_b} is fulfilled, too.

In order to prove a two-photon blockade, it is sufficient to fulfill Eqs.~\ref{Ineq:ImperfectPB_a} and \ref{Ineq:ImperfectPB_b} for $n$=2, which can further be simplified for $\langle\hat m\rangle \ll 1$ to
\begin{subequations}
	\begin{align}
	(i)\quad  & g^{(3)}<  1 \\
	(ii)\quad & g^{(2)}\geq 1.
	\end{align}
\end{subequations}
Note that this derivation was done for a comparison of the output light field with a light field with Poissonian photon number statistics for which all orders of normalized photon correlations are equal to one, which corresponds to the level of uncorrelated photons.

\section*{Apparatus, Atom Trapping, Positioning, and Sequence}
Our system consists of a high-finesse Fabry-Perot resonator with a length of \SI{200}{\micro\metre} and a field decay rate $\kappa/2\pi=\SI{2.0}{\mega\hertz}$ that is loaded with single $^{87}Rb$ atoms (polarization decay rate $\gamma/2\pi = \SI{3.0}{\mega\hertz}$) via an atomic fountain from a magneto-optical trap (MOT) \SI{25}{cm} below (Fig.\,\ref{fig:fig0}).
Atoms passing between the two mirrors are captured in a red-detuned, standing-wave optical dipole trap at \SI{800}{\nano\metre} focussed to the center of the cavity.
After the loading we ramp up another orthogonal transverse, blue-detuned \SI{773}{\nano \metre} standing-wave.
Together with the intracavity trap at \SI{773}{\nano\metre} (\SI{5}{} free spectral ranges detuned from the resonant mode), these form a three-dimensional lattice that confines atoms to far below $\lambda/2$ in all directions\,\cite{Reiserer2013c}.

The required friction force to capture and cool atoms stems from cooling beams counterpropagating along the $\hat{x}$ direction with orthogonal linear polarizations.
This light is red-detuned by about \SI{50}{\mega\hertz} with respect to the $F=2\leftrightarrow F'=3$ transition of the $D_2$ line and leads to intracavity Sisyphus cooling in all directions\,\cite{Nussmann2005a}.
A weak transverse beam resonant to the $F=1\leftrightarrow F'=2$ transition on the $D_1$ line at \SI{795}{\nano\metre} repumps atoms that end up in the $F=1$ ground state due to off-resonant scattering.

The light scattered by the atoms is collected with a high numerical aperture objective ($\text{NA}=\SI{0.47}{}$) and detected using an EMCCD camera with an integration time of \SI{300}{\milli\second}.
From the images, we detect the presence of atoms, their number and position in real-time.
For each captured image, we calculate the deviation of the atom position from the cavity mode center along the $\hat{x}$ direction which is then fed back onto a piezo motor that shifts the standing-wave of the red transverse trap to realign atoms.
During the data analysis, we postselect on single atoms that were well confined to the center of the cavity along $\hat{x}$ and $\hat{z}$.
We have no direct information on the position along $\hat{y}$.
Fluctuations in width of the atom images, however, indicate a significant distribution along this direction on the order of the cavity waist.

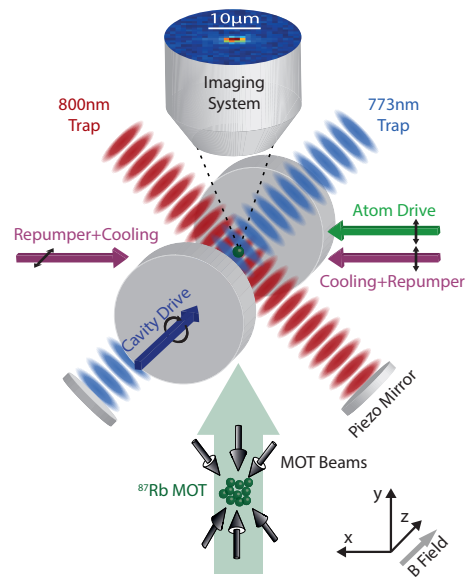
\fig{0}{\sfigZeroFile}{Sketch of the experimental apparatus. For clarity, the intracavity trap is not depicted.}{fig0}
%
The ac-Stark shift, mainly caused by the \SI{800}{\nano\metre} trap, compensates the atom-cavity detuning $\Delta_{ac}/2\pi=(\omega_a-\omega_c)/2\pi=\SI{-15.2}{\mega\hertz}$ to the $F=2\leftrightarrow F'=3$ transition of the $D_{2}$ line at \SI{780}{\nano\metre} to only a few \si{\mega\hertz}.
For the largest dipole matrix element for the Zeeman states $m_F=+2\,\leftrightarrow\,m_F'=+3$ (Fig.\,\ref{fig:fig1}(a)), an atom-cavity coupling constant of $g_0/2\pi=\SI{20}{\mega\hertz}$ puts us well into the strong-coupling regime of cavity QED, $g\gg (\kappa,\gamma)$.

The quantization axis is defined parallel to the cavity axis by an offset magnetic field along the $\hat{z}$ direction of about \SI{0.4}{\gauss}.

Atoms are typically trapped for about \SI{5}{\second}.
During this time, we repeat our measurement sequence as depicted in Fig.\,\ref{fig:fig1}(b) with a rate of \SI{2}{\kilo\hertz} alternating between cavity and atom driving.
We start with a cooling interval of \SI{400}{\micro\second} followed by \SI{50}{\micro\second} of state preparation.
Here, we pump the atom to the $F=2, m_F=+2$ ground state by applying a circularly polarized, resonant cavity probe that drives $\sigma^+$ transitions (light blue arrows in Fig.\,\ref{fig:fig1}(a)).
As the respective Clebsch-Gordan coefficients increase towards the final state, excitation will be increasingly suppressed due to a growing normal mode splitting.
This ensures rapid state preparation with a strong drive while at the same time avoiding excessive heating.
We finish with the probe interval during which either the cavity or the atom drive excites the system at the desired frequency.
We record the transmitted signal on 4 single-photon detectors in a Hanbury Brown and Twiss-type configuration with a timing resolution of \SI{1}{\nano\second}.
Depumping and heating effects are minimized by keeping this interval short (\SIrange{20}{50}{\micro\second}), especially in case of higher driving strengths.
Whereas the cavity probe drives $\sigma^+$ transitions, we excite the atom with a linearly polarized transverse probe that drives $\sigma^+$ and $\sigma^-$ transitions.
Since the dipole matrix element for the $\sigma^-$ transition is much weaker, we expect a quasi two-level behavior.
The power is chosen such that we remain in the weak driving regime, $\eta_{a,c}\leq(\kappa,\gamma)$.
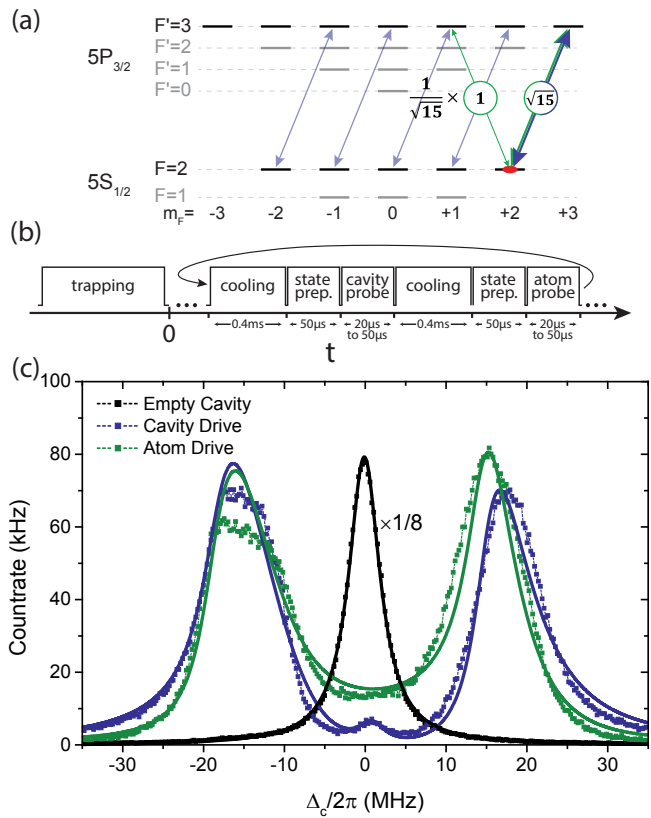
\fig{1}{\sfigOneFile}{(a) Driving scheme for atom (green) and cavity drive (blue). The $\sigma^+$ pump (bright blue) during state preparation populates the $F=2,\, m_F=+2$ ground state. Due to the large difference in Clebsch-Gordan coefficients, the linear transverse probe experiences an almost two-level-system. (b) Experimental sequence used during the experiment. (c) Spectra measured for the uncoupled (black, scaled by factor 1/8) and strongly-coupled (blue) cavity driven system in comparison to exciting the atom (green) (same data set as in Fig. 1(a) of the paper). The thick colored lines are fits of a model considering residual thermal excitation and remnants of the empty cavity for the coupled spectra in case of cavity driving.}{fig1}
%
\section*{Theory Models}
As given in Eq. 1 in the paper, the driven atom-cavity system can be described by the Jaynes-Cummings Hamiltonian\,\cite{Jaynes1963} plus a driving term $H_d$, here in rotating wave approximation:
\begin{equation}\label{Eq.:JC-Hamiltonian}
H=\hbar \Delta_{a}\, \hat{\sigma}^\dagger\hat{\sigma}+\hbar \Delta_{c}\, \hat{a}^\dagger \hat{a} + \hbar g\,(\hat{a}^\dagger \hat{\sigma}+\hat{\sigma}^\dagger \hat{a}) + H_d
\end{equation}
where $\Delta_a=\omega_d-\omega_a$ ($\Delta_c=\omega_d-\omega_c$) is the atom (cavity) detuning with respect to the driving frequency $\omega_d$, $\hat{\sigma}^\dagger$ ($\hat{\sigma}$) is the atomic raising (lowering) operator, and $\hat{a}^\dagger$ ($\hat{a}$) is the photon creation (annihilation) operator with $\hat{n}=\hat{a}^\dagger \hat{a}$ being the photon number operator.

The driving term is either $H_d=\hbar\eta_c\,(\hat{a}+\hat{a}^\dagger)$ when simulating cavity drive or $H_d=\hbar\eta_a\,(\hat{\sigma}+\hat{\sigma}^\dagger)$ when atom driving is investigated.
\subsection*{Semi-classical model and spectra}
Using Eq.\,\ref{Eq.:JC-Hamiltonian} and following the work of \citet{Murr2003a}, it is possible to derive the Heisenberg equations of motion for a set of system operators.
In general these are a set of coupled differential equations which cannot be solved analytically.
Nevertheless, an analytical solution may be found in the regime of low driving when treating the light field classical thus omitting higher excitation rungs. In case of cavity driving we find the following equations of motion:
\begin{subequations}
	\begin{align}
	\langle\dot{a}\rangle = i(\tilde{\Delta}_c\langle \hat{a}\rangle-\eta_c-g\langle \hat{\sigma} \rangle)\\
	\langle\dot{\sigma}\rangle = i(\tilde{\Delta}_a\langle\hat{\sigma}\rangle-g\langle \hat{a} \rangle)
	\end{align}
\end{subequations}
Here, we have used the complex detunings $\tilde{\Delta}_a = \Delta_a+i\gamma$, $\tilde{\Delta}_c = \Delta_c+i\kappa$.
These equations can be solved analytically for the steady state solution of the mean photon number:
\begin{align}
\langle a^\dagger a\rangle_\text{c} = \frac{\eta^2_c|\tilde{\Delta}_a|^2}{|\tilde{\Delta}_c\tilde{\Delta}_a-g^2|^2}
\end{align}
This equation yields the well-known normal mode spectra and includes the antiresonance for cavity driving\,\cite{Sames2014}.
Following a similar approach, the mean photon number can analogously be calculated for the atom driven case:
\begin{align}
\langle a^\dagger a\rangle_\text{a}= \frac{\eta^2_a g^2}{|\tilde{\Delta}_c\tilde{\Delta}_a-g^2|^2}
\end{align}
Due to the light shifts of the optical dipole traps, finite temperature $T$ of the atom causes an inhomogeneous broadening of the atomic resonance which in turn affects the normal mode spectrum.
This can be modeled by averaging over Boltzmann-distributed atomic detunings as described in the Supplementary Information of\citet{Neuzner2016b}.
The model, however, assumes a steady state temperature independent of the probe detuning.
This may lead to deviations since the drive is expected to be our main heating source that strongly depends on its detuning\,\cite{Maunz2005}.

In the cavity driven case, a small peak at the empty cavity resonance frequency can be observed as shown in Fig.\,\ref{fig:fig1}(c), which is identified as empty cavity remnants (about 1\% of the empty cavity transmission).
The origin is an imperfect polarization of the transmitted light as a consequence of cavity birefringence.
To incorporate this technical issue, we extend our previous model by adding another term, a Lorentzian of amplitude $A_{ec}$, width $2\kappa$ and center frequency $\Delta_c$.

In summary, this model is used to fit the spectra in Fig.\,\ref{fig:fig1}(c) corresponding to Fig. 1(e) in the paper using $\left(g, \Delta_{ac}, \eta_{a,c}, T, A_{ec}\right)$ as free parameters.
Of these parameters, we use the system's key parameter $g$ for further calculations of quantum correlations.

\subsection*{Full quantum master equation and \\correlation functions}
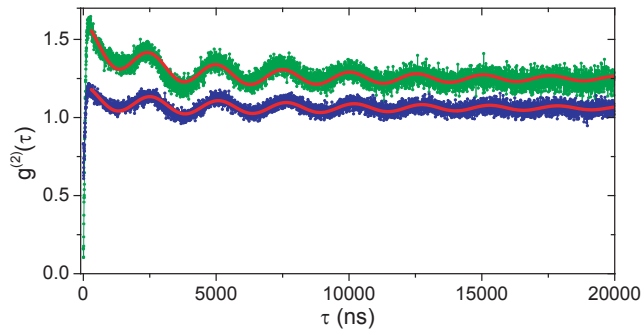
\fig{2}{\sfigTwoFile}{The second-order photon correlation for atom (green) and cavity (blue) driving at the first manifold on a large range of $\tau$ is depicted. From the fit of an oscillating function we deduce a frequency of about \SI{395}{\kilo\hertz} in good agreement with the expectation for our axial trap frequency. In addition, we find a finite offset from one even after decay of the trap dynamics for $\tau > \SI{15}{\micro\second}$.}{fig2}
%
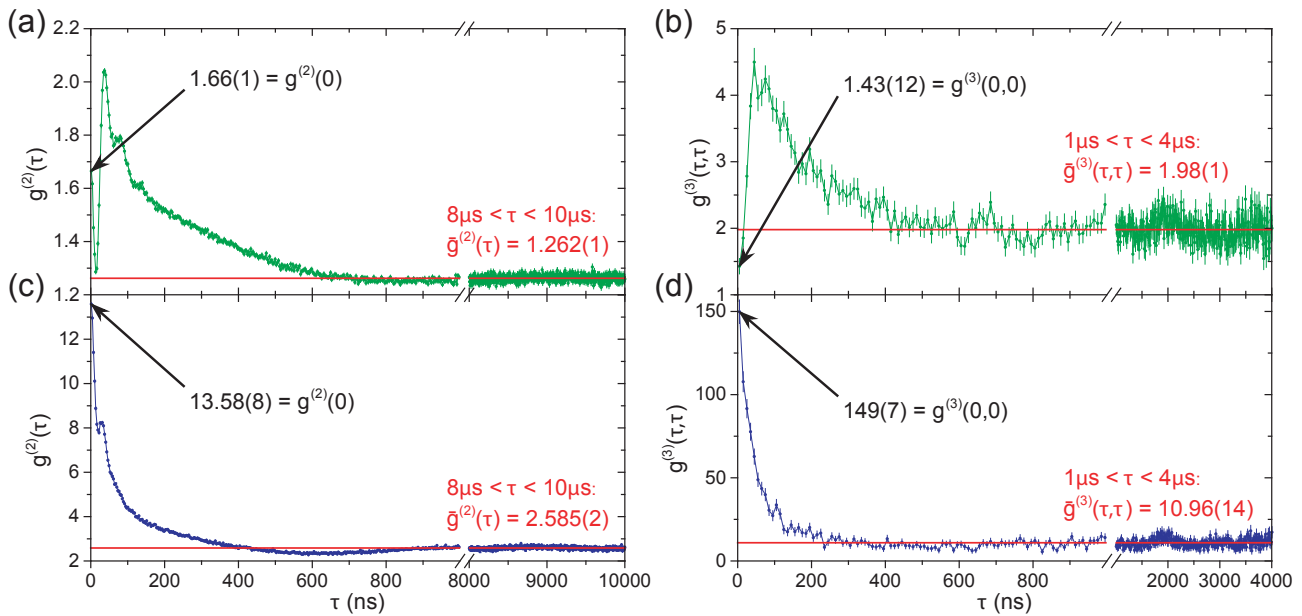
\figx{3}{\sfigThreeFile}{The second- and third-order photon correlations for atom (green, (a) and (b), respectively) and cavity (blue, (c) and (d), respectively) driving close to the second manifold are depicted including their averaged values for large $\tau$. In black we indicate the values for $\tau=$\SI{0}{}. The values $\bar{g}^{(2)}(\tau)$ and $\bar{g}^{(3)}(\tau)$ given in red refer to the averaged values of $g^{(2)}(\tau)$ and $g^{(3)}(\tau)$ over an interval of $\tau$ as indicated.}{fig3}
%
It is mandatory to use the full quantum model in order to simulate effects that rely on the quantum dynamics of the system such as $n$th-order correlations of single- or multi-photon blockade.
As described in textbooks, e.g.\,\cite{Carmichael1993}, we calculate the dynamics of the system using the density matrix ($\rho$) formalism and solve the standard Lindblad Master equation:
\begin{align}\label{Eq.:LindbladME}
\dot{\rho(t)} =& -\frac{i}{\hbar}\left[H,\rho(t)\right] \nonumber \\
&+\sum_{i=1}^{2} \left(2C_i\rho(t)C^\dagger_i-\rho(t)C^\dagger_iC_i - C^\dagger_iC_i\rho(t)\right)
\end{align}
Here, the first term describes the coherent evolution of the system described by the Hamiltonian of Eq.\,\ref{Eq.:JC-Hamiltonian}.
The remaining terms take the system's interaction with the environment into account.
For our open quantum system, this is given by the decay operators $C_{1} = \sqrt{\gamma} \hat{\sigma}$ and $C_{2} = \sqrt{\kappa} \hat{a}$ that describe dissipation of the atom polarization and the cavity field, respectively.

Equation\,\ref{Eq.:LindbladME} can be written as:
\begin{equation}
\dot{\rho}=\mathcal{L}\rho
\end{equation}
with $\mathcal{L}$ being a Lindblad superoperator.
A formal solution is given by:
\begin{equation}
\rho(t)=e^{\mathcal{L}t}\rho(0)
\end{equation}
Therefore, the time dependence of an arbitrary operator $\hat{O}$ is given by:
\begin{equation}
\langle\hat{O(t)}\rangle=\text{tr}(\hat{O}e^{\mathcal{L}t}\rho(0))
\end{equation}
According to the quantum regression theorem the second order correlation function can now be calculated via\,\cite{gardiner2004quantum}:
\begin{equation}
\langle a^\dagger a^\dagger(\tau)a(\tau)a\rangle = \text{tr}(a^\dagger a e^{\mathcal{L}\tau}(a \rho_{ss} a^\dagger))
\end{equation}
where $\rho_{ss}$ is the steady state density matrix of the system. This can be interpreted as suddenly moving the system away from equilibrium by annihilation of a photon and projecting on the new density matrix $a\rho_{ss} a^\dagger$.
The time dependent second order correlation function is then given by the expectation value of the photon number $\langle a^\dagger a\rangle$ during the subsequent equilibration.
This concept can directly be transferred to the third order correlation function and leads to
\begin{align}
&\langle a^\dagger a^\dagger(\tau_1) a^\dagger(\tau_1+\tau_2) a(\tau_1+\tau_2) a(\tau_1)a\rangle \nonumber \\ &= \text{tr}\left(a^\dagger a e^{\mathcal{L}\tau_2}\left[a e^{\mathcal{L}\tau_1}(a \rho_{ss} a^\dagger)a^\dagger\right]\right).
\end{align}

The photon distribution in the cavity mode (as shown in Fig.\,3(c) \& (d) in the paper) can be calculated by tracing out the atomic part of the full steady state system density matrix $\rho_{ss}$.
The diagonal elements of the reduced density matrix are the population of the respective Fock states.

Note that we are considering two different cases here, atom driving and cavity driving.
As was shown in~\cite{Alsing1991}, one can formally transfer the solution of one of these cases into the solution of the other case by applying a simple displacement on the density matrix.
Such a displacement adds photons for the cavity drive compared to the atom drive, which gives an intuitive explanation why the blockade is more difficult to achieve with cavity driving.

The numerical simulations of the system as described in this section are performed using the \textit{Quantum Toolbox in Python} (QuTiP)~\cite{Johansson2013}.
We use the coupling constant from the fits of the spectra, the experimental parameters $\kappa$, $\gamma$, and $\eta_{a,c}$, and calculate the correlations and photon distributions near either the first (Fig. 2) or second manifold (Fig. 3-5) for zero atom-cavity detuning.

\section*{Correlations, Statistics, and Technical Fluctuations}
In theory, second- and third-order photon correlations settle to \SI{1}{} after quantum coherence is lost.
This is typically reached after several excited state lifetimes  ($\sim\left(\frac{\kappa+\gamma}{2}\right)^{-1}=\SI{64}{\nano\second}$).
In contrast, in the experiment the correlation functions settle to values above \SI{1}{}.
Two main reasons are given in the text: position distribution of atoms with respect to the cavity mode center or excitation beam and residual atomic motion.
As stated above, the position distribution is caused by the loading and trapping scheme for atoms while motion results from finite temperatures after cooling and heating due to the resonant probe beams.
Both can be observed in correlations.
In Fig.\,\ref{fig:fig2} we show the second-order correlations on the first manifold for atom and cavity driving up to \SI{20}{\micro \second}.
As demonstrated in the paper, we observe a pronounced antibunching feature at $\tau=0$ and quantum dynamics of the system that settle within hundreds of nanoseconds to a peak resulting from technical fluctuations.
The major contribution of this bunching peak stems from an oscillation of about \SI{395}{\kilo\hertz} that agrees well with twice the atomic trap frequency along the cavity axis determined by the blue-detuned intracavity dipole trap\,\cite{Diedrich1987a,Rotter2008}.
The transverse trap frequencies are on the same order, but only excursions along the standing wave of the cavity will cause a significant modulation of the coupling constant which in turn will cause a breathing of the normal modes.
Consequently, the shape and amplitude of the resulting intensity modulation strongly depend on the driving frequency.
The correlations presented in the paper are taken close to the maximum or at the inside slope of the normal modes and are therefore prone to show strong motion artifacts.
In addition, motion along all trap axes leads to varying atom-cavity detunings.
The motion changes the overlap with the optical dipole traps which in turn affects the corresponding light shift from the dynamical Stark effect.
We expect the resulting background to be incoherent as the variations happen at different frequencies and lack phase stability relative to each other.

In total, we model the long-time behavior of the two-photon correlations by the following function:
\begin{equation}\label{eq:corrfit}
f(\tau)=A_{c}\cdot e^{-\frac{\tau}{\tau_{c}}}\cdot\sin{\left(2\pi f_{c}\tau\right)}+A_{i}\cdot e^{-\frac{\tau}{\tau_{i}}}+c
\end{equation}
While the first two terms describe the intensity fluctuation due to coherent variation of the coupling constant along the cavity axis and incoherent background due to light shifts, the third term describes an offset expected to be \SI{1}{} for a Poissonian light field.
We find values of \SI{1.25}{} and \SI{1.06}{} for atom and cavity drive at the first manifold, respectively.
We attribute this additional offset after decay of all dynamics of the system to the random position of the atoms with respect to the cavity mode center which varies from run to run.
This position distribution causes fluctuations in the coupling constant, light shift, and overlap with the excitation beams.
Therefore, it may be thought of as excess intensity noise of the field emitted from the cavity.
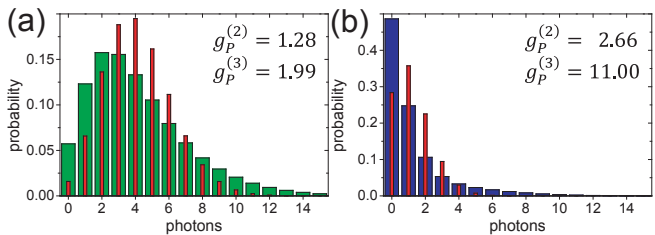
\fig{4}{\sfigFourFile}{Photon distribution $P(n)$ per probe interval for (a) atom and (b) cavity excitation. These distributions can be used to calculate the expected values for a second- and third-order photon correlation of a given light field. The deviation in average photon number is partly caused by different driving strengths ($\eta_a/2\pi\approx\SI{1.6}{\mega\hertz}$ and $\eta_c/2\pi\approx\SI{1.1}{\mega\hertz}$) but mostly due to the different spectral intensities at the insides of the normal modes. The red bars show a Poisson distribution of the same mean photon number.}{fig4}
%

To further investigate the deviation from one, we compare the large-delay offsets of second- and third-order photon correlations for atom and cavity driving close to the second manifold (Fig.\,\ref{fig:fig3}) with the photon number distributions $P(n)$ of the corresponding probing intervals (Fig.\,\ref{fig:fig4}).
The width of the latter already indicates an increased variance in comparison to a Poissonian light field shown as red bars.
Furthermore, we calculate the values for $g^{(2)}$ and $g^{(3)}$ expected from the photon distributions via\,\cite{Rundquist2014}:
\begin{align}
g^{(2)}_P = \frac{\sum_n n(n-1)P(n)}{\left[\sum_n nP(n)\right]^2},\\
g^{(3)}_P = \frac{\sum_n n(n-1)(n-2)P(n)}{\left[\sum_n nP(n)\right]^3}
\end{align}
The resulting values are shown as an inset in Fig.\,\ref{fig:fig4}.
We find good agreement with the large-delay averages $\bar{g}^{(2)}(\tau)$ and $\bar{g}^{(3)}(\tau)$ given in Fig.\,\ref{fig:fig3}.
Apart from the small standard errors in the large-delay averages, we expect a systematic error stemming from the random choice of the averaging interval with respect to residual oscillations due to atomic motion.

This strengthens the claim that correlations do not settle to one as a consequence of excess intensity noise likely caused by the atom position fluctuations.
The cause may be confirmed by investigating the timescale of this behavior which goes beyond the scope of this paper.

In summary, we show that the long-term dynamics exhibit coherent and incoherent behavior.
While the former is assigned to modulation of the coupling strength due to motion along the cavity axis, resulting in coherent oscillations, the latter is compatible with intensity variations due to varying light shifts along all axes, resulting in an exponential decay of the correlations.
The remnant offset at times beyond these dynamics can be assigned to position fluctuations from run to run which lead to an emitted light field that is more chaotic than one following Poisson statistics and therefore has a $g^{(2)}$- and $g^{(3)}$-value for uncorrelated photons exceeding \SI{1}{}.

\end{document}